\documentstyle{elsart}
\begin{document}

\begin{frontmatter}
\title{Varying-$\alpha $ Theories and Solutions to the Cosmological Problems}
\author{John D. Barrow$^1$ and Jo\~ao Magueijo$^2$ }
%EndAName
\address{$^1$Astronomy Centre, University of Sussex,\\
Brighton BN1 9QJ, U.K.\\
$^2$Theoretical Physics, Imperial College,\\
Prince Consort Road, London SW7 2BZ, U.K.}
\date{}
\maketitle

\begin{abstract}
If the fine structure constant $\alpha =e^2/(\hbar c)$ were to change, then
a number of interpretations would be possible, attributing this change
either to variations in the electron charge, the dielectric constant of the
vacuum, the speed of light, or Planck's constant. All these variations
should be operationally equivalent and can be related by changes of standard
units. We show how the varying speed of light cosmology recently proposed
can be rephrased as a dielectric vacuum theory, similar to the one proposed
by Bekenstein. The cosmological problems will therefore also be solved in
such a theory.
\end{abstract}
\end{frontmatter} 

\section{Theories with varying constants}

Numerous experiments have attempted to establish whether or not the
traditional fundamental constants of physics are indeed constants. In such
efforts it is important to recognise that one should consider only
dimensionless constants \cite{beck,dicke}. Measurements of dimensional
quantities represent ratios with respect to standard units. In reference
frames moving relative to each other, or at different points in spacetime,
one cannot be sure of the equivalence of these units. Statements of
constancy of dimensional constants must therefore be circular, because they
postulate the correspondence of the standard units used \cite{btip}.
Conversely, any experimental evidence for varying dimensional constants
could always be absorbed into a redefinition of units. Thus attention must
be focussed upon dimensionless ratios of dimensional constants.

Suppose that evidence is found for varying dimensionless constants. Any
theory explaining the phenomenon would necessarily have to make use of
dimensional quantities. It would be a matter of choice as to which
dimensional quantities were taken to vary. Any theory based on a choice to
vary one dimensional constant could always be reformulated as a theory based
on another choice of varying constant. Thus, evidence for time-variation on
the fine structure constant could be accommodated within a theory assuming
constant electron charge, $e$, and varying speed of light, $c$, or one
postulating varying $e$ and constant $c$. The two formulations should be
equivalent as far as dimensionless observations are concerned. However, the
simplest theories based on different choices would not necessarily be the
same.

Evidence has recently been found \cite{webb} that is consistent with a
time-varying fine structure constant $\alpha =e^2/(\hbar c)$. The
observations make use of high-quality Keck Telescope data and a new
theoretical technique to compare quasar spectral lines of in different
multiplets, simultaneously analysing the MgII 2796/2803 doublet and up to
five FeII transitions (FeII 2344, 2374, 2383, 2587, 2600\AA ), from three
different multiplets. This technique improves our observational sensitivity
to changes in $\alpha $ by an order of magnitude. New upper limits on
possible time-variation in $\alpha $ are found at low redshifts ($z<1$) with 
$\Delta \alpha /\alpha =-1.9\pm 0.5\times 10^{-5},$ consistent with
strongest known limits from the Oklo natural reactor \cite{oklo}, but
evidence is found for a variation is detected at high redshift ($z>1$) with $%
\Delta \alpha /\alpha =-1.1\pm 0.4\times 10^{-5}$ . As yet, this cannot be
taken as positive evidence for time variation in the fine structure constant
at high redshift because it cannot be excluded that the observed spectral
variation arises from subtle line blendings. However, these observations and
the new technique they introduce provide a new level of precision in testing
the constancy of constants which is significantly better than direct
laboratory measurement. Since string theories permit the variation of $%
\alpha $, we would also like to be able to place cosmological limits on the
variation of $\alpha $ by examining its consequences in the very early
universe. In order to do this rigorously we need a self-consistent theory
which incorporates varying $\alpha $ into the cosmological evolution
equations.

Changes in $\alpha $ can be interpreted in different ways. A theory of
varying electric charge was first proposed by Bekenstein \cite{bek2}. In
this theory the vacuum may be seen as a dielectric medium effectively
screening the electric charge. A varying speed of light theory (with $\hbar
\propto c$) was proposed by Albrecht and Magueijo \cite{am}. These two
theories correspond to different representations of a varying $\alpha $ in
terms of varying dimensional constants\cite{note}. There should exist a set
of duality transformations between these two representations. The purpose of
this paper is to provide these transformations and explore the
interconnection between these two classes of theory

The minimal varying-$c$ theory is of interest because it offers a means of
solving the so-called cosmological problems: the horizon, flatness,
cosmological constant, entropy, and homogeneity problems. We would like to
identify the varying-$e$ reformulation of this theory. We will find that
this theory is far from the minimal varying-$e$ theory. It is similar to a
Brans-Dicke theory in which the dielectric field $\epsilon =e/e_0$ plays the
role of the inverse of the Gravitational constant $\phi =1/G$. Such a theory
is formally Lorentz invariant, a major improvement over the varying-$c $
formulation.

%Secondly it would be curious to investigate how the minimal changing 
%$e$ theories behave with respect to the cosmological puzzles. 
The plan of the paper is as follows. In Sec.~\ref{vsl} we review the varying
speed of light proposal. In Sec.~\ref{map} we derive the transformation of
units that maps varying $c$ theories into varying-$e$ theories. In Sec.~\ref
{dual} we write down the varying-$e$ dual to the minimal varying speed of
light theory. Then in Sec.~\ref{cosmdual} we derive the cosmological
equations in this theory. The cosmological problems pose dimensionless
questions. Hence they should also be solved in this theory. We spell out the
solution to the flatness problem in Sec.~\ref{flatness}. We conclude with
some remarks on the proposed theory.

For obvious reasons, we do not use units in which $G=c=\hbar =1$.

\section{Varying speed of light theories}

\label{vsl} In varying speed of light (VSL) theories a varying $\alpha $ is
interpreted as $c\propto \hbar \propto \alpha ^{-1/2}$. The electromagnetic
coupling $e$ is constant, Lorentz invariance is broken, and so by
construction there is a preferred frame for the formulation of the physical
laws. In the minimally coupled theory one then simply replaces $c$ by a
field in this preferred frame. Hence, the action is still 
\begin{equation}
S=\int dx^4{\left( \sqrt{-g}{\left( {\frac{\psi (R+2\Lambda )}{16\pi G}}+%
{\cal L}_M\right) }+{\cal L}_\psi \right) }  \label{svsl}
\end{equation}
with $\psi (x^\mu )=c^4$. The dynamical variables are the metric $g_{\mu \nu
}$, any matter field variables contained in ${\cal L}_M$, and the scalar
field $\psi $ itself. The Riemann tensor (and the Ricci scalar) is to be
computed from $g_{\mu \nu }$ at constant $\psi $ in the usual way. This can
only be true in one frame: additional terms in $\partial _\mu \psi $ must be
present in other frames.

Varying the action with respect to the metric and ignoring surface terms
leads to 
\begin{equation}
G_{\mu \nu }-g_{\mu \nu }\Lambda ={\frac{8\pi G}\psi }T_{\mu \nu }.
\end{equation}
Therefore, Einstein's equations do not acquire new terms in the preferred
frame. Minimal coupling at the level of Einstein's equations is at the heart
of the model's ability to solve the cosmological problems. It requires of
any action-principle formulation that ${\cal L}_\psi $ must not contain the
metric explicitly, and so does not contribute to the energy-momentum tensor.

In a cosmological context, the Friedmann metric can still be written as 
\begin{equation}
ds^2=-c^2dt^2+a^2\left[ {\frac{dr^2}{1+Kr^2}}+r^2d\Omega \right] ,
\label{dsvsl}
\end{equation}
and the Einstein's equations are still, 
\begin{eqnarray}
{\left( {\frac{\dot a}a}\right) }^2 &=&{\frac{8\pi G}3}\rho -{\frac{Kc^2}{a^2%
}}  \label{fried1} \\
{\frac{\ddot a}a} &=&-{\frac{4\pi G}3}{\left( \rho +3{\frac p{c^2}}\right) }
\label{fried2}
\end{eqnarray}
However, the conservation equation that follows from (\ref{fried1})-(\ref
{fried2}) is now 
\begin{equation}
\dot \rho +3{\frac{\dot a}a}{\left( \rho +{\frac p{c^2}}\right) }={\frac{%
3Kc^2}{4\pi Ga^2}}{\frac{\dot c}c}  \label{cons1}
\end{equation}

The definition of ${\cal L}_\psi $ controls the dynamics of $\psi $. It is
only required that $\psi $ does not couple to the metric. In \cite{am} $c$
changes in an abrupt phase transition, but one could also imagine $c\propto
a^n$ as in \cite{jb}. The latter scenario would result from a Lagrangian of
the Brans-Dicke type, with 
\begin{equation}
{\cal L}_\psi ={\frac{-\omega }{16\pi G\psi }}\dot \psi ^2
\end{equation}
(where $\omega $ is a dimensionless coupling constant) and is being
investigated \cite{bmbd}. The addition of an appropriate
temperature-dependent potential $V(\psi )$ could induce a phase transition,
as required in the scenario developed in \cite{am}. In this respect the
abrupt scenario of Albrecht and Magueijo \cite{am} and the smooth scenario
of Barrow \cite{jb}are the analogues of inflationary cosmological evolution
with and without a phase transition, respectively.

\section{Mapping varying-$c$ theories into varying-$e$ theories}

\label{map} Given a variable $\alpha $, and a VSL theory, it must always be
possible to redefine units so that $c$ and $\hbar $ are constant, and $e$
varies. The two descriptions should be equivalent with respect of
dimensionless quantities. To find the transformation let us assume that
measurements of intervals of length $dx$, time $dt$, and energy $dE$, are
made in the VSL system of units. In this system $c\propto \hbar \propto 1/{%
\sqrt{\alpha }}$ and $e=e_0$. Now define a new system of units such that the
same measurements in the new system of units lead to results $d{\hat x}$, $d{%
\hat t}$, and $d{\hat E}$ such that 
\begin{eqnarray}
c_0d{\hat x} &=&c\,dx \\
c_0^2d{\hat t} &=&c^2\,dt \\
{\frac{d{\hat E}}{c_0^3}} &=&{\frac{dE}{c^3}}
\end{eqnarray}
where $c_0$ is a constant, to be identified with the fixed speed of light.
These relations fully specify the new system of units. One may then
construct dimensionless ratios in order to identify the constants in the new
system: 
\begin{eqnarray}
{\frac{{\hat c}d{\hat t}}{d{\hat x}}} &=&{\frac{c\,dt}{dx}}  \label{trans0}
\\
{\frac{{\hat \hbar }}{d{\hat E}d{\hat t}}} &=&{\frac \hbar {dE\,dt}} \\
{\frac{{\hat G}d{\hat E}}{d{\hat x}{\hat c}^4}} &=&{\frac{G\,dE}{dx\,c^4}} \\
{\frac{{\hat e}^2}{d{\hat E}d{\hat x}}} &=&{\frac{e_0}{dE\,dx}.}
\end{eqnarray}
>From these we find that in the new system of units 
\begin{eqnarray}
{\hat c} &=&c_0 \\
{\hat \hbar } &=&\hbar {\frac{c_0}c}=\hbar _0 \\
{\hat G} &=&G \\
{\hat e} &=&e_0{\frac{c_0}c}\propto {\sqrt{\alpha }.}
\end{eqnarray}
Hence, in the new system, $c$ and $\hbar $ are constants and $e$ varies as $%
\sqrt{\alpha }$. The transformation (\ref{trans0}) can also be defined by 
\begin{eqnarray}
d{\hat x} &=&dx/\epsilon  \label{trans} \\
d{\hat t} &=&dt/\epsilon ^2 \\
d{\hat E} &=&dE\epsilon ^3
\end{eqnarray}
where $\epsilon ={\hat e}/e_0$ is the dielectric constant of the vacuum in
the varying-$e$ system.

The transformation (\ref{trans}) between units of energy, length, and time,
determines the relations between any measurements in the two systems of
units. For instance, of mass ${\hat M}=M\epsilon $, of mass density ${\hat
\rho }=\rho \epsilon ^4$, and of pressure ${\hat p}=p\epsilon ^4$. The dual
to minimal VSL therefore does not predict $M\propto \epsilon ^2$ as in
Bekenstein's theory \cite{bek2}, a first indication that the minimal
theories in the two formulations are not equivalent. Notice that $d{\hat t}%
/dt>0$, so that the arrow of time is not reversed by this transformation.
Also notice that although the Compton wavelengths of all particles are
adiabatic invariants in VSL theories, they decay as $\epsilon$ increases in
the changing-$e$ dual of VSL (as indeed in Bekenstein's theory).

\section{The dual of minimal VSL}

\label{dual} Action (\ref{svsl}) is defined in a preferred frame where one
postulates that it includes no terms in the derivatives of $\psi $. In a
cosmological setting this frame is defined by the proper time and the
conformal space coordinates which ensure that $K=0,\pm 1$. In general, this
frame would be defined by a 4 vector $u^\mu $, and the metric could be
written as 
\begin{equation}
g_{\mu \nu }=u_\mu u_\nu +h_{\mu \nu }
\end{equation}
with $h_{\mu \nu }u^\mu =0$ and $u_\mu ^\mu =1$. In the new units one may
choose $u^\mu $ to be the same, and the new metric to be 
\begin{equation}
{\hat g}_{\mu \nu }=u_\mu u_\nu +h_{\mu \nu }/\epsilon ^2.
\end{equation}
This is a statement of the preferred VSL frame.

Under the transformation of units (\ref{trans}) the action (\ref{svsl})
becomes 
\begin{equation}
S=\int d{\hat x}^4{\left( \sqrt{-{g}}\epsilon ^{-2}{\left( {\frac{c_0^4({%
\hat R}+2{\hat \Lambda })}{16\pi G}}+{\hat {{\cal L}}}_M\right) }+{\cal L}%
_\epsilon \right) ,}  \label{se}
\end{equation}
where the curvature is to be computed from the old metric in the usual way.
However, the variation is to be performed with respect to the new metric.
Therefore, the new action, apart from the $\epsilon ^{-2}$ factor, is just
the standard Einstein-Hilbert action to which a ``spatial'' conformal
transformation has been applied. This is reminiscent of Brans-Dicke theory,
which is Einstein's gravity subject to a conformal transformation dependent
on a field $\phi $ which represents the inverse of the gravitational
'constant'.

The ADM formalism may be used to write this Lagrangian more explicitly as 
\begin{equation}
S=\int d{\hat x}^4\sqrt{-{\hat g}}\epsilon {\left[ {\frac{c_0^4}{16\pi G}}{%
\left( ^{(3)}{\hat R}{\left( {\hat h_{\mu \nu }}\epsilon ^2\right) }+\kappa
_{ij}\kappa ^{ij}-\kappa ^2+2{\hat \Lambda }\right) }+{\hat {{\cal L}}}%
_M\right] }+{\cal L}_\epsilon ,  \label{sdual}
\end{equation}
where $h_{\mu \nu }$ and $\kappa _{\mu \nu }=h_\mu ^\alpha h_\nu ^\beta
\nabla _\alpha u_\beta $ are the first and second fundamental forms, and $%
^{(3)}R$ is the Ricci scalar derived from the spatial metric $h_{\mu \nu }$.

The Einstein equations derived from this action are more simply written
using the Hamiltonian formalism, with a 3+1 split induced by vector $u^\mu $
(see for instance \cite{grav}). With a unit lapse and zero shift (temporal
gauge), the Hamiltonian density in Einstein's theory takes the form 
\begin{equation}
{\cal H}=h^{1/2}{\left[ -^{(3)}R+h^{-1}{\left( \Pi ^{\mu \nu }\Pi _{\mu \nu
}-{\frac 12}\Pi ^2\right) }\right] }
\end{equation}
The second fundamental form is given by 
\begin{equation}
\kappa _{\mu \nu }={\frac{{\dot h_{\mu \nu }}}2}
\end{equation}
and the momenta conjugate to the $h_{\mu \nu }$ are given by 
\begin{equation}
\Pi _{\mu \nu }={\frac{\partial {\cal L}}{\partial {\dot h}^{\mu \nu }}}%
=h^{1/2}(\kappa _{\mu \nu }-\kappa h_{\mu \nu }).
\end{equation}
The Hamiltonian constraint is ${\cal H}=0$ and the momentum constraint is 
\begin{equation}
\nabla _\mu (h^{-1/2}\Pi ^{\mu \nu })=0
\end{equation}
The dynamical equations are 
\begin{equation}
{\dot h}_{\mu \nu }={\frac{\delta {\cal H}}{\delta \Pi ^{\mu \nu }}}%
=2h^{-1/2}{\left( \Pi _{\mu \nu }-{\frac 12}h_{\mu \nu }\Pi \right) }
\end{equation}
and 
\begin{eqnarray}
{\dot \Pi _{\mu \nu }}=-{\frac{\delta {\cal H}}{\delta h^{\mu \nu }}}
&=&-h^{1/2}  \nonumber \\
&&{\left( ^{(3)}R_{\mu \nu }-{\frac 12}h_{\mu \nu }^{(3)}R\right) } \\
&&\ +{\frac 12}h^{-1/2}h_{\mu \nu }{\left( \Pi _{\alpha \beta }\Pi ^{\alpha
\beta }-{\frac 12}\Pi ^2\right) }  \nonumber \\
&&\ -2h^{-1/2}{\left( \Pi _\mu ^\alpha \Pi _{\alpha \nu }-{\frac 12}\Pi \Pi
_{\mu \nu }\right) }
\end{eqnarray}
These are Einstein's equations in vacuum. Addition of matter is
straightforward. If one performs the transformation 
\begin{equation}
h_{\mu \nu }\rightarrow h_{\mu \nu }\epsilon ^2
\end{equation}
upon these equations one obtains Einstein's equations, in Hamiltonian form,
in the dual of VSL theory.

%Gauss equation reads (cite whoever was the source of Part III notes) 
%\begin{equation} 
%R=^{(3)}R -2u^\mu u^\nu R_{\mu\nu} +\kappa^2-\kappa_{\mu\nu} 
%\kappa^{\mu\nu} 
%\end{equation} 
%where $\kappa_{\mu\nu}=h_\mu^\alpha h_\nu^\beta \nabla_\alpha u_\beta$ 
%(the so-called second fundamental form; $h_{\mu\nu}$ is the first) 
%and $^{(3)}R$ is the Ricci scalar derived from the projected Rieman 
%tensor. Hence if $\kappa_{\alpha\beta}=0$ the action is  
%\begin{equation}\label{se1} 
%S=\int d{\hat x}^4 
%{\left( \sqrt{-{g}}\epsilon ^{-2}{\left( {c_0^4 \over 16\pi G} 
%(^{(3)}{\hat R}\epsilon^2 -2u^\mu u^\nu {\hat R}_{\mu\nu} 
%2{\hat \Lambda}) 
% +{\hat {\cal L}}_M\right)} +{\cal L}_\epsilon \right)} 
%\end{equation} 
%[EXERCISE: explicitly find the transformation laws for the Ricci  
%scalar. Also find the Einsteins equations in such a theory.] 

The varying-$e$ dual of VSL is therefore not the minimal varying-$e$ theory
(encoded in Postulate P8 of Bekenstein's theory \cite{bek2} which assumes
that the Einstein field equations are left unchanged by the introduction of
varying $e$). Instead, it resembles a Brans-Dicke theory of $\epsilon $, in
which a conformal transformation dependent on $\epsilon $ is applied to
standard gravity. However, this transformation is only applied to the
spatial sections defined by a vector field $u^\mu $. The Lagrangian does not
break Lorentz invariance explicitly, but of course the presence of vector $%
u^\mu $ does. In this respect this theory resembles the Lagrangian written
by Coleman and Glashow \cite{cg}.

\section{Cosmology with a non-minimal varying-$e$ theory}

\label{cosmdual} When applied to cosmology, the transformation (\ref{trans})
changes the line element (\ref{dsvsl}) into 
\begin{equation}
d{\hat s}^2=ds^2/\epsilon ^2=-c_0^2d{\hat t}^2+{\hat a}^2\left[ {\frac{d{%
\hat r}^2}{1+{\hat K}{\hat r}^2}}+{\hat r}^2d{\hat \Omega }\right]
\label{dse}
\end{equation}
with ${\hat a}=a/\epsilon $, and ${\hat K}=K$. Note that after the
transformation of units one must perform a spatial coordinate transformation
so that ${\hat K}=K=\{0,\pm 1\}$. The Friedmann equations in the new units
are therefore 
\begin{eqnarray}
\left( {\frac{{\hat a}^{\prime }}{{\hat a}}}+{\frac{\epsilon ^{\prime }}%
\epsilon }\right) ^2 &=&{\frac{8\pi G}3}{\hat \rho }-{\frac{Kc_0^2}{{\hat a}%
^2},}  \label{fried2p} \\
{\frac{{\hat a}^{\prime \prime }}{{\hat a}}}+{\frac{\epsilon ^{,,}}\epsilon }%
-2{\left( \frac{\epsilon ^{\prime }}\epsilon \right) }^2 &=&-{\frac{4\pi G}3}%
({\hat \rho }+3{\hat p}/c_0^2),
\end{eqnarray}
with a prime denoting $d/d{\hat t}$. One can derive these equations by
applying the transformation (\ref{trans}) to the VSL Friedmann equations, or
by writing the Einstein's equations associated with the action (\ref{sdual})
for line element (\ref{dse}).

The conservation equation in this theory becomes 
\begin{equation}
{\hat \rho }^{\prime }+3{\frac{{\hat a}^{\prime }}{{\hat a}}}{\left( {\hat
\rho }+{\frac{{\hat p}}{c_0^2}}\right) }={\frac{\epsilon ^{\prime }}\epsilon 
}{\left( {\hat \rho }-3{\frac{{\hat p}}{c_0^2}}\right) }-{\frac{3Kc_0^2}{%
4\pi G{\hat a}^2}}{\frac{\epsilon ^{\prime }}\epsilon }  \label{cons2}
\end{equation}
The first source term on the right-hand side is zero in the
radiation-dominated era. The second term on the right-hand side couples to
spatial curvature in the same way as in the VSL theory (see eq. (\ref{cons1}%
) above). In this non-minimal varying-$e$ theory the dielectric properties
of the vacuum induce violations of energy conservation in curved space
times. The coupling is such that if ${\epsilon ^{\prime }/\epsilon }>0$,
only the $K=0$ universe is stable. Energy is produced for subcritical
densities, and taken away for supercritical densities. One should therefore
expect a solution to the flatness problem.

One can derive simple radiation-dominated (${\hat p}={\hat \rho }c_0^2/3$)
solutions to these equations, by noting that $a\propto t^{1/2}$, and so 
\begin{equation}
{\hat a}={\frac 1\epsilon }{\left( \int_0^{\hat t}d{\hat t}^{\prime
}\,\epsilon ^2\right) }^{1/2}
\end{equation}
For a sudden phase transition at time ${\hat t}={\hat t}_c$ in which $%
\epsilon $ jumps from $\epsilon _{-}$ to $\epsilon _{+}$ one has 
\begin{eqnarray}
{\hat a}&&={\hat t}^{1/2} \qquad {\rm for}\qquad {\hat t}<{\hat t}_c 
\nonumber \\
&&={\sqrt{{\left( \frac{\epsilon _{-}}{\epsilon _{+}}\right) }^2{\hat t}_c+{%
\hat t}-{\hat t}_c}} \qquad {\rm for}\qquad {\hat t}>{\hat t}_c
\end{eqnarray}
If $\epsilon _{-}\ll \epsilon _{+}$ then effectively 
\begin{eqnarray}
{\hat a}&&={\hat t}^{1/2} \qquad {\rm for}\qquad {\hat t}<{\hat t}_c 
\nonumber \\
&&={\sqrt{{\hat t}-{\hat t}_c}} \qquad {\rm for}\qquad {\hat t}>{\hat t}_c
\end{eqnarray}
and there is a second big bang at $t_c$. The expansion factor ${\hat a}$
drops to zero at ${\hat t}_c$, and evolution restarts as if $t_c$ had been
the big bang.

Scenarios in which $c\propto a^m$ were considered in \cite{jb}. If $-1<m<0$
(a scenario in which the quasi-flatness problem is solved \cite{bmbd}), then 
${\hat a}={\hat t}^{1/2}$, and we have a normal radiation-dominated universe
expansion factor. However, if the flatness problem is to be solved in a
radiation-dominated universe then we require $m<-1$. Since ${\hat a}\propto
a^{m+1}$, again we have that the expansion factor decreases while $\epsilon $
varies. This type of variation should therefore only occur during a short
period in the very early universe. In both cases we see that
radiation-dominated universes in which the flatness problem is solved
display a decrease in the expansion factor ${\hat a}$, and we have deflation
in these coordinates. Deflation is the natural cosmological setting for a
decreasing $\alpha $ theory. In such scenarios the Compton wavelengths of
all particles decay in time. The unusual coupling to gravity that exists in
this theory ensures that the universe deflates, so that the ratio of Compton
wavelengths to the Hubble length does not decrease faster than in the
standard theory.

\section{The flatness problem}

\label{flatness} We now spell out how the flatness problem can be solved in
these cosmologies. In standard VSL cosmology if we assume that 
\begin{eqnarray}
c &=&c_0a^m  \label{see} \\
p &=&(\gamma -1)\rho c^2  \label{gam}
\end{eqnarray}
then, as we found in \cite{jb}, the flatness problem is solved as $%
a\rightarrow \infty $ if 
\begin{equation}
2m\leq 2-3\gamma  \label{flat}
\end{equation}
which reduces to the standard inflationary condition if $m=0.$ This is just
the condition that the ratio between the energy and the curvature
contributions to expansion evolves as 
\[
F_{VSLT}\equiv \frac \rho {Kc_0^2/a^2}\propto a^{2-3\gamma -2m}\rightarrow
\infty 
\]
for $a\rightarrow \infty $. In deriving this expression we used the solution
for $\rho $ that comes from integrating the conservation equation (\ref
{cons1}) after substituting (\ref{see})-(\ref{gam}).

Now, in the dual theory with constant $c$ and varying $e$, let us assume 
\begin{eqnarray*}
\epsilon &=&\epsilon _0\hat a^n \\
\hat p &=&(\gamma -1)\hat \rho \hat c^2
\end{eqnarray*}
and let us study a similar ratio: 
\[
\hat F_{dual}\equiv \frac{\hat \rho }{Kc_0^2/\hat a^2}\ \ 
\]
Again, we can integrate when $K=0$ to find $\hat \rho =B\hat a^{4n-3\gamma
(n+1)}$. This is what must be approached asymptotically when $K\neq 0$ for
the flatness problem to be solved. We then note that $\hat a=a\epsilon ^{-1}$
so that $a\propto \hat a^{n+1}$. Therefore $a\rightarrow \infty $
corresponds to $\hat a\rightarrow \infty $ iff $n>-1$. Also, since $\epsilon
\propto c^{-1}$ in the dual theory, we have the relations 
\[
\epsilon \propto \hat a^n\propto c^{-1}\propto a^{-m}\propto \hat
a^{-m(n+1)} 
\]
and we see that the constants $n$ and $m$ that we have introduced in the two
frames are related by 
\begin{equation}
n+1=\frac 1{m+1}  \label{rel}
\end{equation}
and therefore 
\[
F_{dual}\propto \hat \rho \hat a^2\propto \hat a^{4n+2-3\gamma (n+1)} 
\]
Hence in the dual theory the flatness problem is solved as $\hat
a\rightarrow \infty $ (which corresponds to $a\rightarrow \infty $ when $%
n+1>0$) if 
\begin{equation}
\ 4n+2-3\gamma (n+1)>0.  \label{c1}
\end{equation}
If we have $n+1<0$ then $a\rightarrow \infty $ corresponds to $\hat
a\rightarrow 0$ and the condition for solving the flatness problem in this
limit is 
\begin{equation}
4n+2-3\gamma (n+1)<0.  \label{c2}
\end{equation}
We can show that {\it in both cases} the conditions on $n$ , (\ref{c1})-(\ref
{c2}) for the resolution of the flatness problem transform, using (\ref{rel}%
) into the condition (\ref{flat}) for its resolution in the VSL theory.

More generally we note that under transformation (\ref{trans}) 
\begin{equation}
{\hat \delta }_\Omega ={\frac{{\hat \rho }-{\hat \rho }_c}{{\hat \rho }_c}}={%
\frac{\rho -\rho _c}{\rho _c}}=\delta _\Omega
\end{equation}
Deviations from critical density are therefore the same regardless of the
system of units, as expected, since the flatness problem is a dimensionless
question. The varying-$e$ duals of VSL scenarios which solve the flatness
problem must therefore solve this problem also.

>From Eqns.\ref{fried2} and \ref{cons2} one can derive 
\begin{equation}
{\hat \delta }_\Omega ^{\prime }=(1+{\hat \delta }_\Omega ){\hat \delta }%
_\Omega {\left( {\frac{{\hat a}^{\prime }}{{\hat a}}}+{\frac{\epsilon
^{\prime }}\epsilon }\right) }(3\gamma -2)-2{\frac{\epsilon ^{\prime }}%
\epsilon }{\hat \delta }_\Omega
\end{equation}
%For $\gamma=4/3$ (radiation) 
%there are no direct terms in $\epsilon '/\epsilon$. 
%One still has ${\hat \delta}_\Omega\propto {\hat a}^2$.  
%However now energy density drives the sum of ${\hat a}'/{\hat a}$  
%and $\epsilon '/\epsilon$. An increasing $\epsilon$ forces the Universe 
%to deflate thereby pushing ${\hat \delta}_\Omega$ to zero. 
To first order in ${\hat \delta }_\Omega $ we therefore have 
\begin{equation}
{\hat \delta }_\Omega ^{\prime }={\hat \delta }_\Omega (3\gamma -2){\sqrt{%
\frac{8\pi G\rho }3}}-2{\frac{\epsilon ^{\prime }}\epsilon }{\hat \delta }%
_\Omega
\end{equation}
For a sufficiently fast increase in $\epsilon $ one will therefore always
have ${\hat \delta }_\Omega ^{\prime }/{\hat \delta }_\Omega \ll -1$, and
the flatness problem is solved.

\section{Conclusions}

Theories of varying $\alpha $ have been proposed in the past\cite{bek2},
attributing this change to a change in $e$. These theories couple minimally
to gravity and therefore do not solve the cosmological problems. The VSL
proposal is a varying-$\alpha $ theory which attributes this change to $%
\hbar $ and $c$ instead. VSL theories that are minimally coupled to gravity
can solve the cosmological problems. In this paper we stressed the existence
of a duality between varying-$e$ theories and VSL theories. We derived the
standard unit transformation linking these two types of theory and derived
the dual of minimal VSL. The resulting varying-$e$ theory is of Brans-Dicke
type where the vacuum dielectric field $\epsilon =e/e_0$ behaves like the
inverse of a gravitational coupling $\phi =1/G$. Standard Brans-Dicke theory
may be thought of as a $\phi $-related conformal transformation applied to
Einstein's gravity. The VSL dual derives from an $\epsilon $-related
conformal transformation that only acts on spatial sections of the metric,
as defined by a given vector $u^\mu $. The presence of this vector in the
Lagrangian breaks Lorentz invariance, as in the theory of Coleman and
Glashow \cite{cg}. The resulting theory is comparable to an ether theory in
the sense that the dielectric medium is not just another cosmic ingredient
(as is the case for Bekenstein's theory \cite{bek2}) but also participates
in the formulation of the laws of physics, and defines a preferred frame. We
have showed how in this theory, in scenarios in which the flatness problem
is solved, one has ``deflation''. The increasing $\epsilon $ field induces a
period of contraction of the universe, in accord with the decreasing Compton
wavelengths of all particles.

{\bf Acknowledgments} JDB is supported by the PPARC and JM by the Royal
Society.


\begin{thebibliography}{99}
\bibitem{note}  We are assuming that structure constants $\alpha
_i=e_i^2/(\hbar c)$ based on interactions other than electromagnetic would
also change, keeping fixed ratios with $\alpha $.
\end{thebibliography}
\end{document}